\documentclass[twocolumn,showpacs,preprintnumbers,amsmath,amssymb]{revtex4}

\usepackage{graphicx}
\usepackage{dcolumn}
\usepackage{bm}

\begin{document}

\title{A new synchronization mechanism via Turing-like microscopic
   structures \\
for CO oxidation on Pt(110) }

\author{R. Salazar}
\author{A.P.J. Jansen}
\affiliation{
Schuit Institute of Catalysis (ST/SKA), Eindhoven University of
Technology, P.O. Box 513, 5600 MB Eindhoven, The Netherlands.}

\author{V.N. Kuzovkov}
\affiliation{
Institute for Solid State Physics, University of Latvia,
Kengaraga 8, LV-1063, Riga, Latvia.}

\date{\today}

\begin{abstract}
  We discuss an alternative to the traditional gas-phase coupling
  approach in order to explain synchronized global oscillations in CO
  oxidation on Pt(110). We use a minimalist microscopic model which
  includes structural Pt surface reconstruction via front propagation,
  and large diffusion rates for CO.  The synchronization mechanism is
  associated with the formation of a Turing-like structure of the
  substrate.  By using large parallel microscopic simulations we derive
  a scaling laws which allow us to extrapolate to realistic diffusion
  rates, pattern size, and oscillation periods.
\end{abstract}



\maketitle


Oscillations appear in experiments of CO oxidation on Pt$(110)$ in a
very narrow parameter interval \cite{imb1995}. Additionally, it is also
possible to observe pattern formation; about $T \sim 500$K a crossover
exist from spirals and fronts at low $T$ to standing waves and chemical
turbulence at high $T$. Experiments using fast scanning tunneling
microscopy and field ion microscopy have clearly shown that fast kinetic
processes are typically accompanied by the appearance of microscopic
structures \cite{win2002,vol1999,win1997}.

Although the diffusion of CO is fast, it does not provide a
synchronization mechanism for the global oscillations, it is thought
that the mechanism for global synchronization is coupling via the
gas-phase \cite{lev1992}. This mechanism has been used widely in
classical reaction-diffusion (RD) models. However, experiments have
shown that particle dynamics on surfaces can be highly correlated
\cite{win2002}, something which is not taken into account in RD models.
This leads to differences between RD and experiments which can only be
removed by including phenomenological corrections.  An alternative
mechanism for global synchronization of oscillation has been suggested
based on spontaneous phase nucleation \cite{kor1999,kor2002}. This nucleation
results in a random creation of dynamic defects and leads to global
synchronization via stochastic resonance.

We present in this paper Monte Carlo (MC) simulations of a microscopic
model for CO oxidation on Pt$(110)$. We show that our model can produce
global oscillations via a new mechanism showing microscopic structures
without gas-phase coupling, coupling via stochastic resonance, or
lateral interactions.  The microscopic structures appear in the
substrate, but they also support pattern formation in the adlayer on a
mesoscopic length scale.  In addition we provide a connection between
these MC simulations and experimental results, via scaling laws.


MC simulations provide a systematic approach to include microscopic
effects \cite{luk1998}.  MC simulates directly the chemical reaction
steps in a model, without averaging out particle correlations.
Simulations of our model were performed by using a Cellular Automaton
(CA) which has been shown to be equivalent to MC \cite{kor1998}.  A
limitation of MC simulations has been the inability to deal with
experimental system sizes and realistic diffusion coefficients. However,
by using large parallel simulations as described in \cite{sal2002} we
can now estimate results for experimental conditions.  Using this
parallel code we are able to simulate system sizes of about $1\,\mu$m
with diffusion coefficients of about $10^{-10}$cm$^2$sec$^{-1}$.  This
approaches experimental conditions closer than any other previous MC
simulation.  Experimental values are for system sizes $10^2$ $\mu$m or
more and diffusion coefficient 10$^{-6}$cm$^2$sec$^{-1}$.  There is a
gap between experimental and simulated rate constants, but, as we can
vary the system size and the diffusion rate over a large range, we can
derive scaling laws, which we can use to extrapolate the behavior
predicted by our model to realistic system sizes and diffusion rates.

Our model has already been used in several studies
\cite{kor1999,kor1998b,kuz1999}.  O$_2$ adsorbs dissociatively onto two
nearest neighbor sites with rate constant $(1-y)s_{\chi}$ with $\chi =
\alpha$,$\beta$, where $\alpha$ denotes the $1 \times 2$ phase and
$\beta$ denotes the unreconstructed $1 \times 1$ phase of Pt(110)
\cite{imb1995}.  The experimental value for the ratio of the sticking
coefficients of O$_2$ on the two phases is $s_{\alpha}:s_{\beta} \approx
0.5:1$. CO is able to adsorb onto a free surface site with rate constant
$y$ and it desorbs from the surface with rate constant $k$. Both
reactions are independent of the surface phase to which the site
belongs. In addition CO is able to diffuse via hopping onto a vacant
nearest neighbor site with rate constant $D$. The CO$+$O$\to$CO$_2$
reaction occurs with rate constant $R$, when CO and O are nearest
neighbors sites.  CO$_2$ desorbs immediately forming two vacant sites.
The $\alpha \rightleftharpoons \beta$ surface phase transition is
modeled as a front propagation with rate constant $V$. For two nearest
neighbor surface sites in the state $\alpha \beta$ the transition
$\alpha \beta \to \alpha \alpha$ ($\alpha \beta \to \beta \beta$) occurs
if none (at least one) of these two sites is occupied by CO.
Summarizing the above transition definitions written in the more usual
form of reaction equations gives:
\begin{eqnarray*}
&& \mbox{CO(g)} + S^{\chi} \rightleftharpoons \mbox{CO(a)}, \\
&& \mbox{O$_2$(g)} + 2S^{\alpha} \to 2\mbox{O(a)}, \\
&& \mbox{O$_2$(g)} + 2S^{\beta} \to 2\mbox{O(a)}, \\
&& \mbox{CO(a)} + S^{\chi} \to S^{\chi} + \mbox{CO(a)}, \\
&& \mbox{CO(a)} + \mbox{O(a)} \to \mbox{CO$_2$(g)} + 2S^{\chi}, \\
&& S^{\alpha} \rightleftharpoons S^{\beta},
\end{eqnarray*}
where $S$ stands for a vacant adsorption site, $\chi$ stands for either
$\alpha$ or $\beta$, and (a) and (g) for a particle adsorbed on the
surface or in the gas phase, respectively. For additional details see
Ref.~\cite{kor1999b}.  Our simulations a use special normalization of
the rates: the adsorption rate for CO is related to the partial
pressures by
\begin{equation}\label{y1}
y=\frac{P_{CO}}{P_{CO}+P_{O_2}} .
\end{equation}
Comparing our rate constants with typical experimental values used in RD
models \cite{bar1994,oer2000} we can estimate the unit of time in our
simulation to be about $t_0 \approx 10^{-2} s$. All our rate constants
are expressed in that unit of time. Lengths are expressed in units of
the unit cell parameter $a$.  The diffusion rate $D$ is related to the
experimental diffusion coefficient $D_A$ by $D_A=a^2 D/ z t_0$, where
$z=4$ is the coordination number.


The quality of the oscillations as a function of $y$ has a typical
resonance form. Well-defined oscillations are found near $y=0.494$.
Under certain condition, which we discuss below, these
oscillations are global. Moving away from $y=0.494$ we see first that
the synchronization decreases and then a disappearance of the
oscillations. We restrict ourselves to $y=0.494$ (for a fixed values
$k=0.1$ and $V=1$) in our simulations.

We have found that two situations are possible depending the system
size, the diffusion rate, and on the initial conditions (IC).  If the
system is small or the diffusion is fast we find global oscillations. If
the diffusion is too slow for a given system size the IC become
important. We can again have global oscillations or the formation of
patterns in the form of moving fronts and spirals.

\begin{figure}
\includegraphics[width=\hsize]{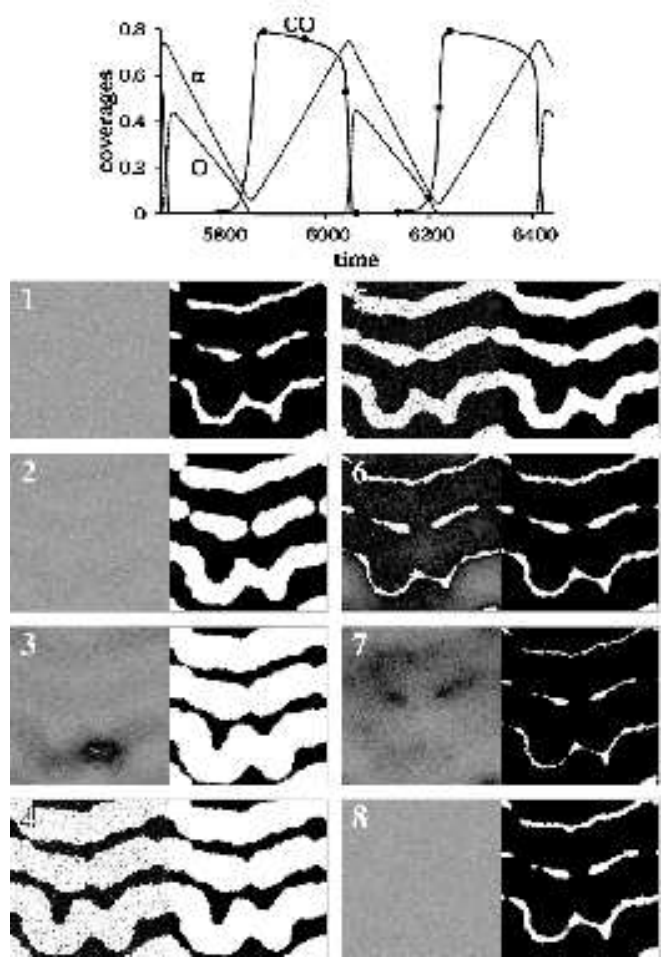}
\caption{ \label{fig:largeD} Oscillations for $D=4000$ and $L=1024$ on the
  coverage CO (solid), oxygen (dashed), and $\alpha$ (dotted). Sequence
  of temporal snapshots correspond to the points in the upper figure.
  Each snapshot has two parts: In the left part we plot the chemical
  species; CO particles are grey and O particles are white, and empty
  sites are black. The right part shows the structure of the surface;
  $\alpha$ phase sites are black, and $\beta$ phase sites are white. }
\end{figure}

To illustrate the synchronization mechanism Fig.~\ref{fig:largeD} shows
global oscillations.  When the system is covered almost fully by CO the
$\beta$ islands in the $\alpha$ background grow (snapshots 1 and 2).
When the size of the $\beta$ islands is large enough the rate of
adsorption of O$_2$ becomes larger than the adsorption of CO on the
$\beta$ zones (snapshots 3 and 4). At the same time the CO in the
$\alpha$ zones is converted to CO$_2$ near to the borders with the
$\beta$ zones (snapshot 4). When the total coverage of oxygen becomes
larger than CO, then the direction of the island grow is reversed
(snapshots 4, 5, and 6) and $\alpha$ islands grow in the $\beta$
background.  When the size of the $\beta$ zones is not large enough to
keep the oxygen coverage larger than the CO coverage (snapshot 7), then
both phases are become covered by CO, and the cycle start again
(snapshot 8). The key point in this cycle is the existence of a critical
size of the $\beta$ islands where oxygen becomes more stable than CO and
a critical size of $\alpha$ islands where CO is more stable than oxygen.
This corresponds to a phase transition produced by a varying size of
these islands. This phase transition is driven by diffusion and the
critical size of the islands depends on the diffusion rate as
$\sqrt{D}$.

\begin{figure}
\includegraphics[width=\hsize]{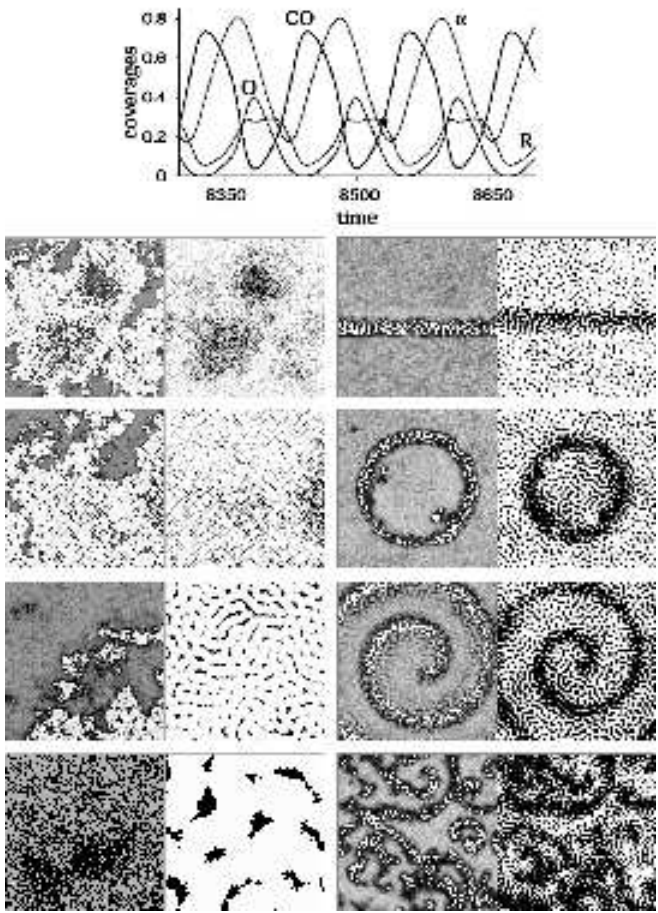}
\caption{ \label{fig:smallD} Global oscillations and pattern
  formation with $D=250$. Sections of the upper-left corner with
  $L=8192$, $4096$, $1024$, and $256$ are shown on the left side. The
  sections correspond to the dot in the temporal plot at the top. It
  shows the same information as Fig.~\ref{fig:largeD}, but it also
  includes the rate of ${\rm CO}_2$ formation $R$.  On the right side we
  have a wave front, a target, a spiral, and turbulence ($L=2048$),
  which can be obtained with different initial conditions.}
\end{figure}

We found that the most important feature of the structures of the
substrate is the typical distance between the islands; i.e., the
correlation length $l_c$ of these structures. For $l_c$ we use the
distance to the first maximum in the radially averaged correlation
function of the substrate.  The shape of the islands self-organizes
through successive oscillations so that they cover the whole system with
same sized islands separated by similar distances and small differences
in size are removed.  We note that our simple model of surface
reconstruction by border propagation dynamic is sufficient to produce
this effect. It yields synchronization at large scales even with
relatively small diffusion rates. In Fig.~\ref{fig:largeD} the diffusion
length $\xi=\sqrt{DT}$, with $T$ the period of the oscillations, is
larger than the system size $L$. The synchronization is therefore
trivial.  In Fig.~\ref{fig:smallD}, however, a simulation is shown where
the diffusion length is much smaller than the lattice size, but larger
than the correlation length. So the diffusion synchronizes the
oscillations on neighboring islands, and the self-organization leads to
global oscillations.

In cases where the synchronization is even better, as with large $D$
values, islands tend to be grouped forming rolls which are parallel (see
Fig.~\ref{fig:largeD}). The structures formed in this way resemble
labyrinthic Turing structures. Moreover a percolating connection between
islands can be observed. The full analytical analysis of this behavior
is far from trivial, however the difference of the mobility for CO and
O, could be sufficient for get a Turing instability in this model.

When the diffusion is slow pattern formation can also occur as on the
right of Fig.~\ref{fig:smallD}.  The distance between the centers of
$\alpha$ and $\beta$ islands is the same as for the global oscillations,
but the size of the islands differ for different parts of the system.
These differences will not be removed as in the fully synchronized
cases. As a consequence a delay or detuning will appear in the
oscillations for different parts of the surface, which can create
fronts, target pattern patterns, spirals, and turbulence. A
cross-section of one of these fronts shows a delay between the pulse of
the adsorbates and the pulse of the substrate phases, which determines
the direction of propagation.  This is a common feature of excitable,
bistable systems, such as the activator-inhibitor models. The most
remarkable feature is that these fronts appear on top of fixed
Turing-like structures which are present even in these incompletely
synchronized cases. So we have within our model two different length
scales for pattern formation; excitable dynamics for fronts and spirals
on a mesoscopic scale (Hopf instability, \cite{bar1994}), and
quasi-fixed Turing-like microscopic structure.

\begin{figure}
\includegraphics[width=\hsize]{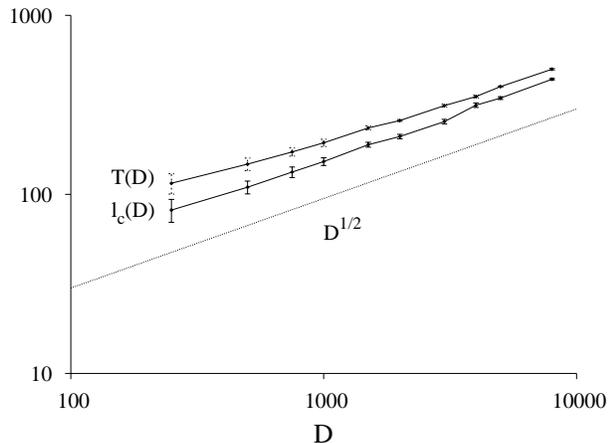}
\caption{ \label{fig:scaling} Scaling behavior for the period of
   oscillations $T(D)$ and the correlation length $l_c(D)$ (units). }
\end{figure}

A scaling analysis for the Turing structures and the oscillations for
the fully synchronized case is presented in Fig.~\ref{fig:scaling}.  We
show the scaling behavior for the period of oscillations $T$ and the
correlation length $l_c$ as a function of diffusion rate $D$.
Numerically we found that the relations $T(D) \sim D^\nu$ and $l_c(D)
\sim D^\mu$ holds with $\nu=0.4849 \pm 0.016$ and $\mu=0.5069 \pm 0.01$.
For large $D$ values a large system size is required to compute $l_c(D)$
precisely. For instance for $D=8000$ we use $L=8192$. Additionally large
diffusion rates means that most of the time is spend moving particles;
on $8000$ diffusion steps only one chemical reaction happens. Also the
simulation time increase because the oscillation period increases with
$D$.

A relationship $T(D) \sim l_c(D) \sim D^{1/2}$ seems quite reasonable,
as our model has fast diffusion. In fact the scaling law for $l_c$ is
the consequence of scaling law for $T$.  The width of the Turing
structure oscillates on a length of order $l_c$ due to the phase
propagation with rate $V$. The period is related to these by $l_c \sim
VT$.  If the velocity $V$ is constant, then $l_c \sim T$.
Fig.~\ref{fig:scaling} was computed for a fixed value of $V$. Using a
larger value for the velocity parameter we can obtain larger correlation
lengths $l_c$ and smaller periods $T$, but the scaling remains the same.
We have found for the Turing-synchronized oscillations that the relation
$T \sim V^{-1/2}$ also applies as in \cite{kor1999,kor1999b,kor2002}.
So we have an important scaling law in form $T = c \sqrt{D/V}$, where $c
\sim 10$.  An estimate for the oscillation period for realistic
diffusion rates $D_A\sim 10^{-6}\,\hbox{cm}^2\hbox{sec}^{-1}$ we have
$D\sim 10^8$. The experimental oscillations period is $\tau=t_0T$ with
$t_0\sim 10^{-2}\,$sec and $\tau=10\,$sec. The value $V$ predicted by
our model for that value of $\tau$ is then of the order of $V\sim 10^2$.

We propose that the minimal condition to have global oscillations is
that the diffusion length $\xi$ is at least of the order of the
correlation length $l_c$: $\xi = l_c$ ($\xi=\sqrt{DT}$ and  $l_c = VT$, $c \sim 10$).
Critical values for the diffusion rate and the
correlation length are then
\begin{eqnarray}\label{eq23}
D & = & c^2 V^3 \\ \nonumber
l_c & = & c^2 V^2. \nonumber
\end{eqnarray}
For $V=1$ we get the order of the parameters used in
Fig.~\ref{fig:smallD} (left side), $D\sim 10^2$ and $l_c\sim 10^2$ where
global oscillations are possible but infrequently.

Diffusion is a thermally activated process so we can compare our
proposal, Eqs.~(\ref{eq23}), with the experimental crossover at $T \sim
500$; for lower temperatures where fronts and spirals appear, we have
lower diffusion rates, $\xi < l_c$ and the synchronization mechanism is
not stable and produces patterns similar to on the right of
Fig.~\ref{fig:smallD}.  On the other hand for large temperatures where
standing waves and chemical turbulence appear, we have large diffusion
rates, $\xi > l_c$, and we get better synchronization, as in
Fig.~\ref{fig:largeD}.  For $\xi = l_c$ we get the crossover,
Eqs.~(\ref{eq23}) holds and we get the minimum criteria for
synchronization, as in Fig.~\ref{fig:smallD} on the left.

Using $V=10^2$ in Eqs.~(\ref{eq23}) we get for the critical value of
diffusion rate $D \sim 10^8$, which corresponds to the experimental
value of $D_A\sim 10^{-6}$cm$^2$sec$^{-1}$. This confirms our crossover
idea. For correlation length $l_c \sim 10^6$ (in units of cell
parameter $a$) we have a size $\sim 10^2$ $\mu$m which is of the order
of magnitude of the standing waves observed in experiments
\cite{oer2000}. Near crossover it is possible to interpret $V$ also as
the velocity $v$ of the fronts and spirals. For $V=10^2$ we have $v=a
V/t_0 \sim 10^{-4}$cm s$^{-1}$, in agreements with
experiments\cite{fal1992}.

To summarize we present in this paper a new mechanism for
synchronization of global oscillations based on microscopic
Turing-like structures in the reconstruction of
the surface. In cases with incomplete synchronized oscillations
the adlayer has a mesoscopic second characteristic length.
This is the first actual demonstration of a double length scale.
Scaling laws are analyzed to connect the model to experimental
parameter values.

This work was supported by the Nederlandse Organisatie voor
Wetenschapperlijk Onderzoek (NWO), and the EC Excellence Center of
Advanced Material Research and Technology (contract N
1CA1-CT-2080-7007). We would like to thank the National Research
School Combination Catalysis (NRSCC)
for computational facilities.

\end{document}